\documentclass[conference,compsoc]{IEEEtran}
\usepackage{epsfig}
\usepackage{syntonly}
\usepackage{rotating}
\usepackage{amsmath}
\usepackage{setspace}
 \usepackage{verbatim} 
\usepackage{amssymb}
\usepackage{amsmath}
\usepackage{graphicx}
\usepackage{multirow}
\usepackage{colortbl}
\usepackage{enumerate}
\usepackage{color}
\usepackage{wrapfig}
\usepackage{subfigure}
\usepackage{url}

\usepackage{ifthen}
\usepackage[dotinlabels]{titletoc}
\usepackage{colortbl}
\usepackage{array}

\usepackage{booktabs}
\usepackage{morefloats}
\usepackage{balance}
\usepackage[above]{placeins}

\usepackage{amssymb}
\usepackage{amsmath} 
\usepackage{centernot}
\usepackage{color}
\usepackage{algorithm}
\usepackage{url}
\usepackage{graphicx}
\usepackage{xfrac}
\usepackage{subfigure}
\graphicspath{{fig_pdf/}}
\DeclareGraphicsExtensions{.pdf, .png}

\numberwithin{equation}{section}

\ifCLASSOPTIONcompsoc
  \usepackage[nocompress]{cite}
\else
  \usepackage{cite}
\fi
\ifCLASSINFOpdf
\else
\fi
\begin{document}
%


\title{\vspace{-45pt} CATERPILLAR: Coarse Grain Reconfigurable Architecture 
\\ for Accelerating the Training of Deep Neural Networks \vspace{-20pt}}

\author{\IEEEauthorblockN{
	Yuanfang Li\IEEEauthorrefmark{1}, 
        Ardavan Pedram\IEEEauthorrefmark{1}\IEEEauthorrefmark{2} }

    \IEEEauthorblockA{\IEEEauthorrefmark{1}Stanford University,   \IEEEauthorrefmark{2}Cerebras Systems}
    \{yli03,perdavan\}@stanford.edu}



%


\maketitle



%
\IEEEpeerreviewmaketitle

\begin{abstract}
Accelerating the inference of a trained DNN is a well studied subject. In this paper we switch the focus to the training of DNNs. 
The training phase is compute intensive, demands complicated data communication, and contains multiple levels of data dependencies and parallelism.
This paper presents an algorithm/architecture space exploration of efficient accelerators to achieve better network convergence rates and higher energy efficiency for training DNNs. We further demonstrate that an architecture with hierarchical support for collective communication semantics provides flexibility in training various networks performing both stochastic and batched gradient descent based techniques. Our results suggest that smaller networks favor non-batched techniques while performance for larger networks is higher using batched operations.  At 45nm technology, CATERPILLAR achieves performance efficiencies of 177~GFLOPS/W at over 80\% utilization for SGD training on small networks and 211~GFLOPS/W at over 90\% utilization for pipelined SGD/CP training on larger networks using a total area of 103.2~$\mathbf{mm^2}$ and 178.9~$\mathbf{mm^2}$ respectively.
\end{abstract}

\section{Introduction}

State of the art Deep Neural Networks (DNNs) are becoming deeper and can be applied to a range of sophisticated cognitive tasks such as image recognition~\cite{cnn} and natural language processing~\cite{rnn}. Convolutional Neural Networks (CNNs)~\cite{cnn} and Recurrent Neural Networks (RNNs)~\cite{rnn} are some of the commonly used network architectures that are inspired by the Multilayer Perceptron (MLP)~\cite{mlp}. Most of the community has focused on acceleration of the forward path/inference for DNNs, neglecting the acceleration for training~\cite{SzeSurvey,TPU}. Training DNNs is a performance and energy costly operation that routinely takes weeks or longer on servers~\cite{gpu}. This makes the task of navigating the hyper parameter space for network architecture expensive. However the nature of computation in training DNNs makes it an excellent candidate for specialized acceleration if the necessary computation/communication functionality is supported~\cite{sys_array}. 

Today, acceleration of the training process is primarily performed on GPUs~\cite{gpu}. However, GPUs suffer from fundamental computation, memory, and bandwidth imbalance in their memory hierarchy~\cite{TPU}. Thus, the fundamental question is what are the compute architecture, memory hierarchy, and algorithmic tradeoffs for an accelerator designed to train deep neural networks. In this paper we aim to address the design tradeoffs by introducing a Coarse Grain Reconfigurable Architecture (CGRA) for training MLPs.

We focus on training MLPs, an important class of DNNs currently used on state of the art servers~\cite{TPU}, with several variants of Backpropagation (BP)~\cite{bp} learning. Although the computation and the communication graph for CNNs and RNNs are more complex compared to MLPs, we chose MLPs for the following reasons. First, MLPs are inherently memory bound and more challenging to accelerate~\cite{dark}. Second, several research studies on the principles of BP and optimization in DNNs investigate MLPs because of their simpler network-architecture~\cite{dfa_results,stochastic,hogwild}. Finally, MLPs represent the fully-connected layers of CNNs~\cite{cnn}. Hence, we believe this effort establishes a platform for future work on accelerating the training of CNNs and RNNs.

The challenge in training DNNs is that the fastest converging gradient descent algorithms in terms of passes over the dataset are not efficient in conventional architectures as they are based on GEneral Matrix-Vector multiplication (GEMV), which is a bandwidth limited operation. For sparse network weights, sparse matrix-vector multiplication is even less compute intensive. However, as the size of the network shrinks, and is able to fit on local storage, the overall cost of communication drops~\cite{spmv}.  GPUs use a variant of DNN training that groups samples into batches in order to overcome bandwidth limitations at the cost of more epochs to convergence and perform GEneral Matrix-Matrix multiplication (GEMM), which is a compute intensive kernel. 

Given the fundamental differences in various techniques for training MLPs, we aim to compare the design space of accelerators and the memory hierarchy for various configurations of networks and training algorithms. This paper makes the following contributions:

\begin{figure}[!t]
\vspace{-15pt}
\centering
\includegraphics[width=0.25\textwidth]{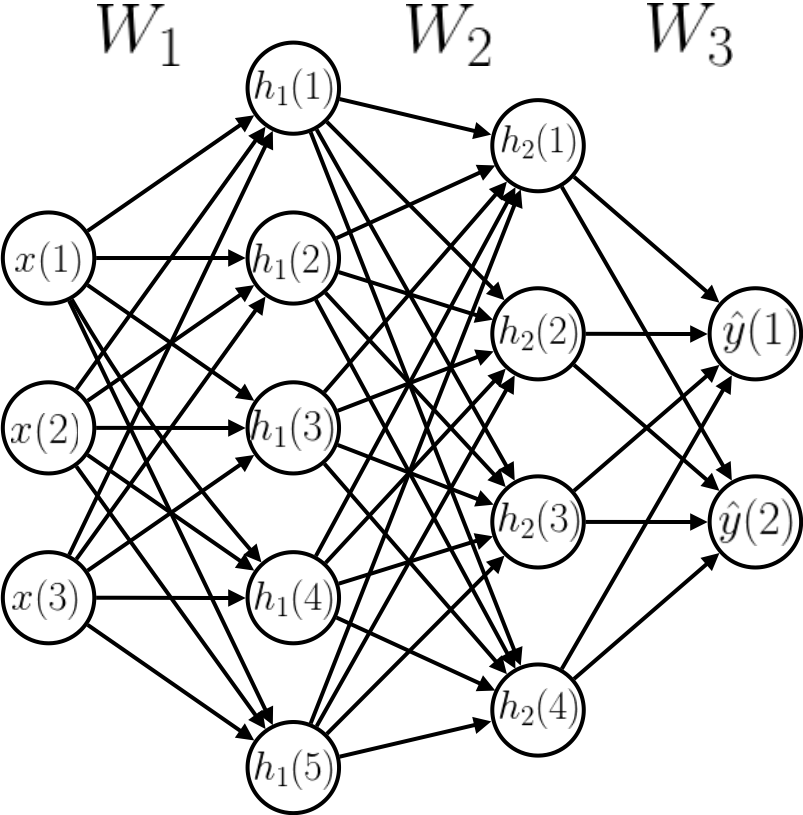}
\vspace{-10pt}
\caption{An MLP with 2 hidden layers $h_1$ and $h_2$.}
\vspace{-15pt}
\label{fig:mlp_diagram}
\end{figure}

\begin{figure*}[!pt]
\centering
\vspace{-20pt}
\includegraphics[width=.8 \textwidth]{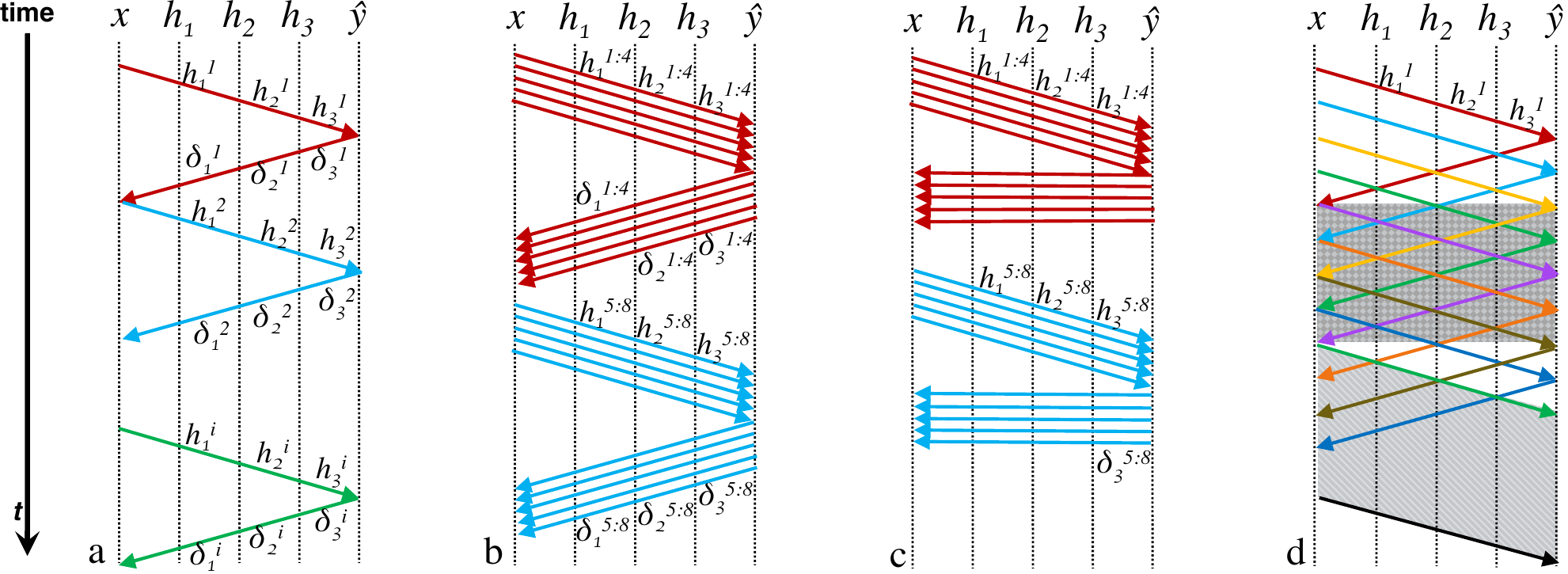}
\vspace{-10pt}
\caption{Methods of learning. Each arrow demonstrates one sample's movement through the forward and backward path in time for a network with three hidden layers. (a) SGD, (b) MBGD, (c) DFA, (d) CP.}
\vspace{-10pt}
\label{fig:learning}
\end{figure*}

\begin {itemize}
\setlength{\itemindent}{.1em}
\item Techniques to decrease dependencies, and increase locality and parallelism for training DNNs and their effect on convergence.
\item Exploration of the design space of accelerators for various BP algorithms.
\item CATERPILLAR: an efficient architecture for training DNNs targeting both GEMV and GEMM kernels exploiting optimized collective communications.
\item Evaluation of the costs of convergence of various networks, training algorithms, and accelerators with respect to both performance and energy efficiency.

\end{itemize}

The rest of the paper is organized as follows: Section~\ref{sec:mlp} describes various BP algorithms for training DNNs.  Section~\ref{sec:architecture} discusses the proposed CGRA, and the algorithm/architecture tradeoffs.  Section~\ref{sec:results} presents the experimental results. Section~\ref{sec:related} reviews related work.  Finally, we conclude in Section~\ref{sec:conclusion}.

 

\section{Multilayer Perceptron~(MLP)} \label{sec:mlp}

The key purpose of the MLP is to output a prediction (generally a classification label) for a given input. Figure~\ref{fig:mlp_diagram} shows a three layered MLP, where $x$ is the input, $\hat{y}$ is the output, and $h_1$ and $h_2$ are the activations of the first and second hidden layers respectively.  $W_i$ is the set of weights such that $W_i(j,k)$ is the weight from the $j$th element of the input to the $k$th element of the output at layer $i$.  At each neuron of a hidden layer a nonlinear activation function $f$ is performed on the sum of the weighted inputs.  At the output layer the softmax function turns $\hat{y}$ into a vector of probabilities for the possible classes.  A bias can be added by appending a +1 term to the input vector at each layer.

The network is trained on input-label pairs $(x,y)$ where $x$ is the input as before and $y$ is the correct label represented as a one-hot vector.  Training is performed in two stages.  During the forward pass the prediction, $\hat{y}$, is calculated:
\begin{align*}
a_1  = x^TW_1, h_1 = f(a_1)
\\a_i = h_i^TW_i, h_{i+1} = f(a_i)
\\a_y = h_2^TW_3, \hat{y} = softmax(a_y)
\end{align*}
During the backward pass, the error in the prediction is calculated at the output and backpropagated to previous layers to provide an estimate of the weight's gradient with respect to the error.  The weights are then updated using gradient descent:
\begin{align*}
e &= \hat{y} - y
\\\delta_2 &= e\odot f'(h_2)W_3^T
\\\delta_1 &= \delta_2\odot f'(h_1)W_2^T
\\W_3 &= W_3 -\eta h_2^Te
\\W_2 &= W_2 - \eta h_1^T\delta_2
\\W_1 &= W_1 - \eta x^T\delta_1
\end{align*}
It is apparent that MLP training consists of a series of large GEMV and GEMM operations, making it an ideal candidate for specialized hardware acceleration. Once the MLP model is trained, inference is performed on an unseen sample by running the forward pass only.
Although MLP is a simpler neural network compared to CNNs and RNNs, by increasing the size and number of the hidden layers as well as the number of training samples, it performs as well as other complex models on tasks like digit classification~\cite{mlp_diff_ref}.



\subsection{Stochastic/Minibatch Gradient Descent (SGD/MBGD)}

In SGD (Figure~\ref{fig:learning}(a)), each training sample goes through the forward and backward passes and updates the weights. Thus, a single pass through a training set of size $K$ results in $K$ updates to the weights~\cite{bp}.  MBGD (Figure~\ref{fig:learning}(b)) is a variant of SGD that groups training samples into "minibatches".  The weights are updated after each minibatch goes through the network so that a single pass through the training set results in $K/b$ weight updates, where $b$ is the size of the minibatch. This allows some parallelization of the training process as several samples can be processed at once. However, the sequential nature of BP still prevents parallelization across layers.

\subsection{Feedback Alignment (FA)}

MLPs are meant to mimic how the brain works, but a BP algorithm such as SGD is not biologically plausible as it requires weight symmetry in the forward and backward pass.  To circumvent this, Lillicrap et al. propose the use of a fixed random feedback weight, $B_i$, for each layer during the backward pass  \cite{dfa_orig}.  The change in learning is:
\begin{align*}
e &= \hat{y} - y
\\\delta_2 &= eB_2^T\odot f'(h_2)
\\\delta_1 &= \delta_2B_1^T\odot f'(h_1)
\end{align*}
Despite the use of random feedback weights, FA can perform as well or better than standard MBGD. However, it is necessary to either use batch normalization or significantly lower the learning rate to prevent the gradients from growing too large, especially for deep networks~\cite{dfa_results}.

\subsection{Direct Feedback Alignment (DFA)}

DFA (Figure~\ref{fig:learning}(c)~\cite{dfa}) backpropagates the last layer's weights to all previous layers as follows:
\begin{align*}
e &= \hat{y} - y
\\\delta_2 &= eB_2^T\odot f'(h_2)
\\\delta_1 &= eB_1^T\odot f'(h_1)
\end{align*}
Like FA, DFA also requires either batch normalization or a smaller learning rate.  However for a typical neural network, the dimension of the output layer is significantly smaller than the hidden layers.  This means that the $B_i$'s will be significantly smaller than their corresponding $W_i$'s, reducing the number of computations required to backpropagate the error by up to several orders of magnitude.  Figure~\ref{fig:learning}(c) shows DFA also introduces parallelism during the backward pass as the error for all layers can be calculated at once.

\subsection{Pipelined/Continuous Propagation ((MB)CP)}
\label{sec:CP}

 Continuous propagation allows parallelization across layers by relaxing the constraint that a full forward and backward pass through the network is required for each set of samples before the weights can be updated\cite{cp_orig}.  In CP, forward and backward passes through a layer can occur simultaneously - in particular the weights can be applied to the current sample on its forward pass at the same time that the previous sample updates the weights on its backward pass.  Figure~\ref{fig:learning}(d) demonstrates the CP algorithm as samples are propagated through the layer and time.  Once the pipeline has been initialized, all layers are simultaneously working.  

\section{CATERPILLAR Architecture} \label{sec:architecture}

\begin{figure*}[!t]
\centering
\vspace{-25pt}
\includegraphics[width=.9 \textwidth]{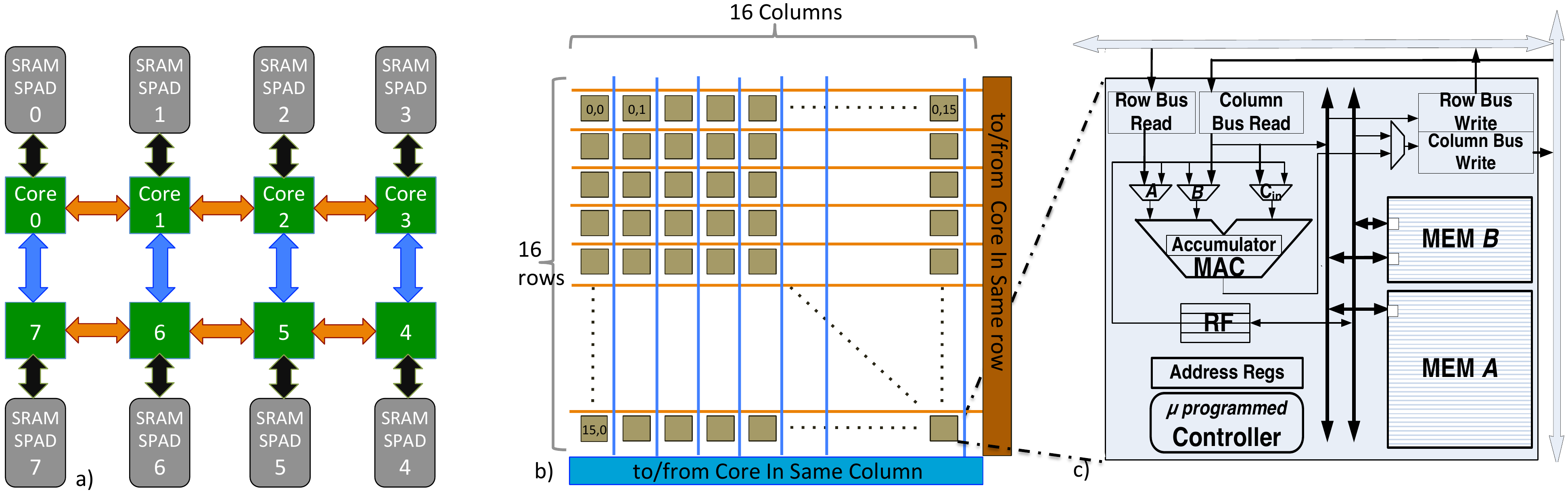}
\vspace{-10pt}
\caption{CATERPILLAR:  (a) Array of cores with ring communication; (b) core with $16\times16$ PEs connected to column and row broadcasts; (c) PE~\cite{lac}.}
\label{fig:architecture}
\vspace{-10pt}
\end{figure*}

This section motivates the design of the CATERPILLAR architecture by demonstrating the available parallelism and locality for each of the existing MLP training algorithms in Section~\ref{sec:mlp}. We then build on these insights and propose an architecture that provides the required computation and communication functionalities.

\subsection{Discussion of Various Algorithms with Respect to Locality and Parallelism}

Training the neural network requires a series of GEMV and/or GEMM operations. There are two cases to consider: (1) the entire network's weights fits in the local memories and (2) it does not fit and some of the weights must be stored off-core. Below we briefly describe various methods of exploiting parallelism and locality in the different training methods.

In the SGD algorithm, only a single sample passes through the entire network at once.  Thus, the compute kernel of this algorithm is GEMV, which is memory bound and inefficient, especially when the entire network does not fit in the aggregate local memories of each core.  For a matrix of size $m\times n$ GEMV requires $mn$ weight accesses to produce $n$ elements of the output.  Thus each of the proposed variants on BP aims to overcome this drawback by introducing and exploiting various sources of parallelism and locality.

Note that even in case (1) when the entire network is stored locally, SGD remains inefficient as GEMV itself is inherently an inefficient operation especially when performed on a 2D array of processing elements. In addition to broadcasting the input vector across one dimension, a reduction must be performed across the opposite dimension to sum up the partial sums into a single output vector.  Consequently, there are $m$ broadcasts and $n$ reductions to transform an input of size $m$ to an output of size $n$.

\textbf{Data parallelization:}  Algorithms that use minibatches (MBGD, MBCP, DFA) exploit parallelism in sample data by passing several samples through the network at once.  This transforms the computations in each layer of the network into a series of GEMM operations.  Furthermore, by accumulating the gradient estimates from several samples into one weight update, the number of memory accesses needed for the weights is divided by the batch size.  The further the minibatch size is increased, the greater parallelism achievable, but if the minibatch size is too large, the algorithm will fail to converge to an acceptable value. In practice, minibatch sizes on nodes range from 2 to 100~\cite{batch_size} and can go up to 10,000+ on clusters~\cite{asgd}.

An important caveat of data parallelization is that for all BP algorithms, the activations calculated during the forward pass must be stored for use during BP.  Thus as the minibatch size grows, more memory is needed to store the activations.

\textbf{Activation locality:}  During the backward pass, the activations, $h_i$, computed during the forward pass are needed in order to update the weights, introducing activation locality between the forward and backward pass.  However, layers are visited in the reverse order during BP so that for a single sample, the first activation to be produced is the last one to be consumed.  For SGD/CP, the size of the activations is negligible compared to the weights, in which case, the activations can be stored without incurring significant memory overhead.  While activation size is still smaller than weight size for algorithms using minibatches, as the network grows deeper, the total size of activations to be stored grows and the memory overhead becomes a concern.  For a network with $L$ layers trained using minibatches of size $b$, if the first layer is $m\times n$, the size of activations to be stored is $Lbn$.  If $Lb\geq n$, this becomes larger than the size of the layer's weights.  This issue is more evident in CNNs than MLPs, as they are usually deeper and have huge sample sizes~\cite{cnn_depth}.  This issue is mitigated with reverse checkpointing and recomputing the activations in earlier layers~\cite{MemeffB}.  Here, only the activations for some layers are saved.  During BP  when an activation that has not been saved is needed, the network propagates the last saved activation for that input through the forward path again.

\textbf{Layer parallelization and weight locality:}  CP introduces layer parallelization by pipelining the samples through the network.  Instead of distributing the weights for a layer to all cores, as in SGD, they can be distributed to a subset of cores, creating a pipeline of layers mapped on the cores and allowing all layers to be processed simultaneously.  With each GEMV operation distributed to a smaller number of PEs, the number of reductions needed decreases, increasing utilization.  Additionally, the ability to perform the forward and backward passes through a layer simultaneously allows the weights updated by the backward pass to be applied immediately to the forward activations, thus decreasing the memory accesses for the weights by half compared to SGD.

\textbf{Dependency elimination:}  By propagating the output layer's error to all previous layers, DFA eliminates the dependency between layers during the backward pass, allowing all weight updates to proceed in parallel.  If the size of the network is large enough that all compute units are busy, this provides little advantage.  However, in the case of small networks, this allows parallelization of the layers during the backward pass.

\subsection{CGRA for Training}

The presented locality and parallelism exploration demonstrates that for networks which do not fit in the local memory of the cores, an architecture optimized for GEMM will perform well by performing a minibatch learning algorithm.  However, the same architecture must also support the high communication demands of GEMV operations if the network is small enough to be stored locally, in which case either SGD or CP can be used to train the network without incurring memory access and communication overheads.

Since GEMM and GEMV inherently use the same inner kernel, an architecture inspired by an array of Linear Algebra Cores (LACs)~\cite{lac,LAP_TC12} will perform well.  The LAC consists of $n_r\times n_r$ Processing Elements (PEs).  Each PE contains a half precision floating point multiply-accumulate unit (FPU) and a local SRAM.  PEs are connected across columns and rows by low-overhead broadcast buses. The architecture uses a $2\times C$ array of these cores connected in a systolic fashion that can support unidirectional ring communication.  Communication between cores is systolic such that the number of cycles to pass data from one core to another is equal to the distance between the cores.  Each core also has its own private off-core memory.  Figure~\ref{fig:architecture} shows the multicore architecture for an array of $2\times 4$ cores with $16\times 16$ PEs each.  The following section presents a detailed description of how different training methods map to the architecture.

\subsection{Mapping of Various Learning Methods}

\textbf{CP/MBCP:}  To perform CP, the architecture must support fast broadcast and reduction of partial products between and within the core for GEMV.  To compute the matrix-vector multiplication within each core, the weights are distributed to the array of PEs in 2D round robin fashion and the input vector to the layer is broadcast across the row buses.  Each PE performs a MAC operation and produces a partial sum of the output vector.  To sum the partial products, each PE broadcasts its value along the column bus and these values are summed together in the diagonal PEs, which broadcasts the final output vector out along the row buses.

\begin{figure}[t]
\vspace{-15pt}
\centering
\includegraphics[width=0.37\textwidth]{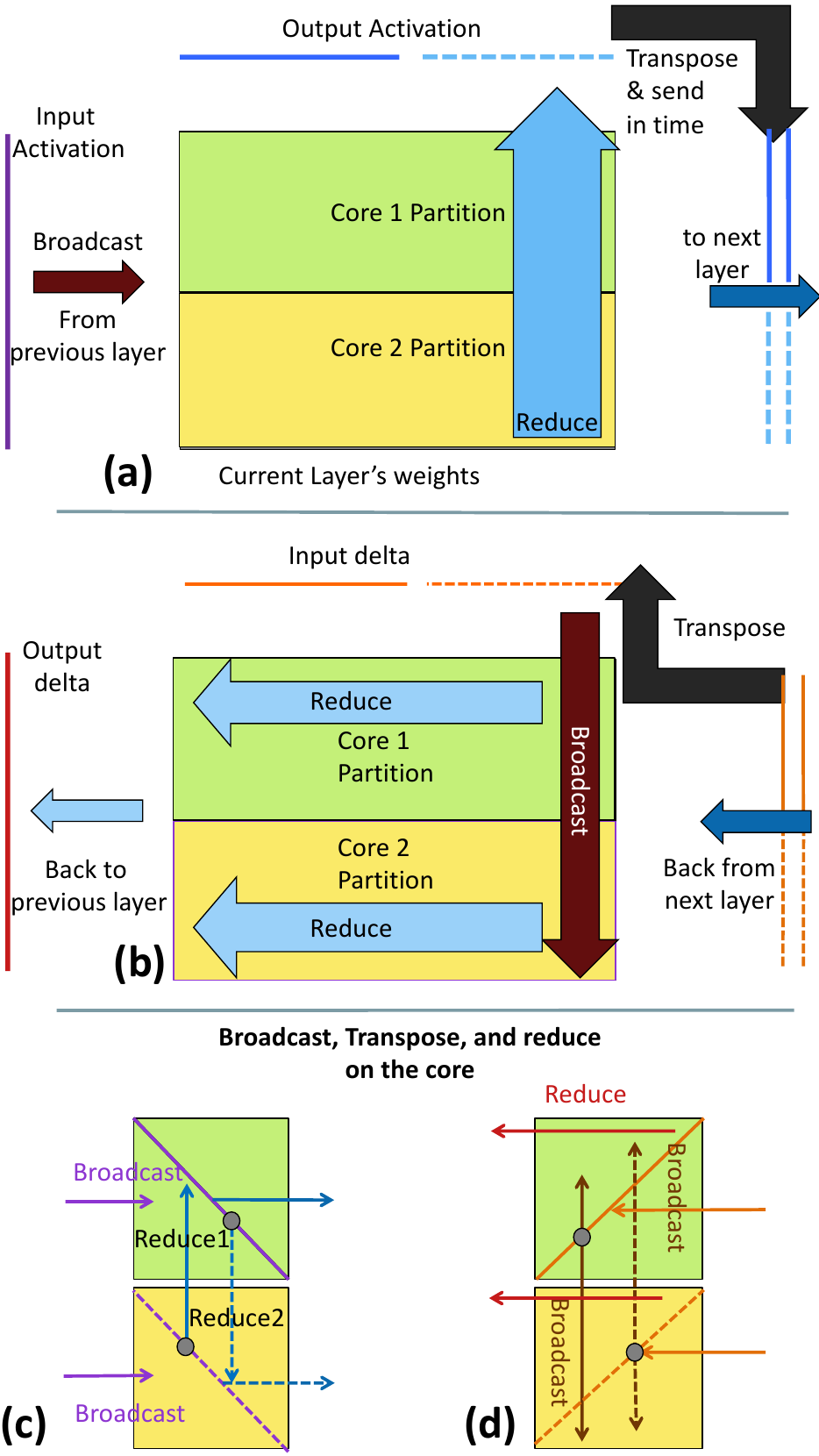}
\vspace{-10pt}
\caption{GEMV across multiple cores:  (a), (c) forward pass through layer; (b), (d) backward pass through layer.}
\vspace{-15pt}
\label{fig:gemv}
\end{figure}

Layers of the network may not be the same size, thus larger layers must be assigned to more cores to keep the cores busy with a stream of activations.  To address the lack of symmetry between the number of cores across layers, a method of reducing partial sums across a non-square array of cores is demonstrated in Figure~\ref{fig:gemv} for the case of two cores exploiting the diagonal PEs in each core.  

In the forward pass (a), each core receives a portion of the input activation and calculates a partial product of the output.  Each core then performs the reduction internally, and subsequently passes its result to the other to sum it to the final output (c).  In the backward pass (b), the input is broadcast across both cores and each core reduces within itself to produce a portion of the output (d).  Note that GEMV produces a transposed output, thus in (a) the reduction must occur in the diagonal PEs and in (b) the diagonal PEs broadcast to produce an untransposed output.

The LAC architecture contains fast broadcast buses.  However, the reduction operation remains expensive as each PE in a column must broadcast its partial sum, with $(n_r-1)$ cycles required to produce one output for an array of $n_r \times n_r$ PEs.  As the reduction is performed on the diagonal PEs, the remaining PEs are idle during this time.  Fortunately, the pipelined nature of CP allows an overlap in computation of the next sample either in the forward or backward pass with reduction of the current sample.  There is no conflict in the broadcast buses, as reductions use the column buses and broadcasts during computation use the row buses (this is reversed for the backward pass).  Thus the only overhead in CP is the extra cycles required to fill and empty the pipeline, which is proportional to the depth of the network.  

\textbf{SGD:}  For the SGD algorithm, since there is limited parallelism between layers, a single GEMV operation is performed by all of the PEs. This means the computation for all layers will be performed sequentially and by the same PEs. The mapping of CP and SGD are therefore similar with one major difference: Instead of passing the result of the current layer to the next set of cores,  it is rebroadcast to perform the GEMV operation for the next layer. 

The expensive overhead of reduction can drop the utilization to half. To address the reduction overhead, direct communication is added between neighboring PEs. This drops the cost of reduction of $n_r$ elements in $n_r$ PEs from $n_r-1$ to $log(n_r)-1$ as it allows more parallel communication between short distance PEs.

\textbf{MBGD:}  For MBGD with small batch sizes, training occurs in the same method as SGD.  During minibatch training with larger batches, all cores are working on the same layer and produce a single matrix output.  Within cores, GEMM occurs as described in~\cite{lac}, but cores must also now be able to pass results to each other between layers.  

In the forward pass, the $i$th core contains a row panel $W_i$, of the weights and accesses all elements of the input activation $X$, to produce a row block $Y_i$ of the output activation.  Below we show an example for three cores:
\begin{align*}
\footnotesize
\begin{bmatrix}
Y_1\\
Y_2\\
Y_3
\end{bmatrix}
=
\begin{bmatrix}
W_1\\
W_2\\
W_3
\end{bmatrix}
\begin{bmatrix}
X
\end{bmatrix}
\end{align*}
In order to make the complete output $Y$ available to all cores as input for the next layer, an all-gather~\cite{collective} operation is performed between layers where each core passes its $Y_i$ to the next, using the ring communication.  For a ring of $2 \times C$ cores with $n_r^2$ PEs per core,$(nb-nb/c)/n_r$ cycles are required to communicate an output of size $n\times b$. 

In the backward pass, the errors need to be multiplied by the transpose of the weights from the forward pass.  As each core's off-core memory contains only a portion of the weights, a transpose is performed with the weights in place by changing the input.  Each core now contains a column panel of the weights $W_i^T$ and receives a row block of the input error, $Y_i^T$, then calculates a partial product of the output error:
\begin{align*}
\footnotesize
\begin{bmatrix}
X_1
\end{bmatrix} 
+ 
\begin{bmatrix}
X_2
\end{bmatrix}
+ 
\begin{bmatrix}
X_3
\end{bmatrix}
=
\begin{bmatrix}
W_1^T & W_2^T & W_3^T \end{bmatrix}
\begin{bmatrix}
Y_1\\
Y_2\\
Y_3\\
\end{bmatrix}
\end{align*}
To obtain the final output a reduce-scatter ~\cite{collective} operation is performed.  The systolic ring communication between cores is used with the same communication overhead as for all-gather.

\textbf{FA:}  FA is best used when memory limitations are not an issue, since twice the weight accesses and storage is required for both $W_i$ and $B_i$ compared to SGD.  In addition, the need for batch normalization makes FA impractical architecturally since the inputs to each layer must either be normalized or the network must be trained for more epochs to achieve similar convergence as MBGD.  Liao et al. also showed that the use of feedback alignment led to no performance improvement over traditional gradient descent when applied to a MLP network trained on TIMIT dataset~\cite{dfa_results}.  Thus feedback alignment is not further considered in the current study on MLPs although it may be revisited for CNNs.

\textbf{DFA:}  The difference between DFA and MBGD is that during the backward pass, error is not propagated between layers, i.e., the output error of the current layer is not needed as input to the next layer.  However the reduce-scatter operation is still required to sum the partial sums to obtain the final output, thus the mapping to the architecture is the same.

\textbf{Activation Function:  }For all algorithms, to calculate the nonlinear activations at each layer, Goldschmidt's method~\cite{nonlinear1}~\cite{nonlinear2}~\cite{nonlinear} is used, which can be implemented with a lookup table and the existing FPU in the PE.  Calculation of the activation thus requires a local memory access to the lookup table and a few iterations of multiply and accumulate operations.  Note that the derivative of the activation required during the backward pass can be easily calculated, as for typical nonlinearities it is a linear function of the activation itself, i.e. for the sigmoid activation function it is 
$\sigma'(x) = \sigma(x)(1-\sigma(x))$.

\subsection{Architectural Tradeoffs}

In a single epoch, each sample passes through the network once on the forward pass and once on backward pass for all BP algorithms, but the number and manner of weight updates varies.  Thus the number of FPU operations required for each algorithm is the same and only the amount of memory accesses, local storage and overhead differs.  The exception is the DFA algorithm, which calculates each hidden layer's error using the last layer's error.  For all other algorithms, each sample has a forward pass, a backward pass and a gradient calculation for all layers, resulting in a total of $3K\sum\limits_{i=1}^Lm_in_i$ MAC operations for a network of size $L$ with $m_i\times n_i$ layers trained on K samples.  For DFA, the backward pass requires only $K\sum\limits_{i=1}^Lm_in_L$ operations as each set of weights is applied to the last layer's error.  The activation function for each layer must also be applied to each sample, but this is negligible compared to the cost of the matrix multiplications.

For SGD and CP, which use GEMV operations, to avoid high communication cost with off-core memory, the network must be able to fit onto the local core memory.  In addition to storing the weights for each layer, the input and activation for the layer must also be stored for use during BP.  When performing the GEMV operation, extra memory is also required to store the partial sums, thus the total memory required for the network to fit is $\sum\limits_{i=1}^L(L-i+1)(m_i+n_i+max(m_i,n_i)+m_in_i)/(2Cn_r^2)$, where $n_r^2$ is the number of PEs/core and $2C$ is the number of cores.  If this is larger than the available memory, part of the network must be stored off-core, which will impact the effective utilization if the off-core memory bandwidth is insufficient.

For MBGD, the weights can be stored off-core and thus local memory is only required to store the activations, which are size $\sum\limits_{i=1}^L(L-i+1)(m_i+n_i)b$ where $b$ is the minibatch size.  If this is larger than the available memory, only part of the activations can be stored and reverse checkpointing must be used.

In SGD, DFA and MBGD, the weights are accessed once during the forward pass and once during the backward pass for each weight update performed.  Thus for a single epoch over $K$ samples, SGD requires $2K\sum\limits_{i=1}^Lm_in_i$ accesses while MBGD requires $(2K/b)\sum\limits_{i=1}^Lm_in_i$ accesses.  DFA also requires an additional $(K/b)\sum\limits_{i=1}^Lm_in_L$ accesses to the random weights $B_i$ during the backward pass.  Additional memory accesses are required to apply the activation function during the forward pass and access the activation during the backward pass but these are on the order of $K\sum\limits_{i=1}^Ln_i$ and thus negligible compared to weight accesses.

The CP and MBCP algorithms allow the weights to be accessed only once for both the forward and backward pass, thus the number of memory accesses is halved compared to SGD/MBGD to $(K/b)\sum\limits_{i=1}^Lm_in_i$.

\begin{figure*}[!t]
\centering
\vspace{-25pt}
\begin{tabular}{cc}
\includegraphics[width=0.48\textwidth]{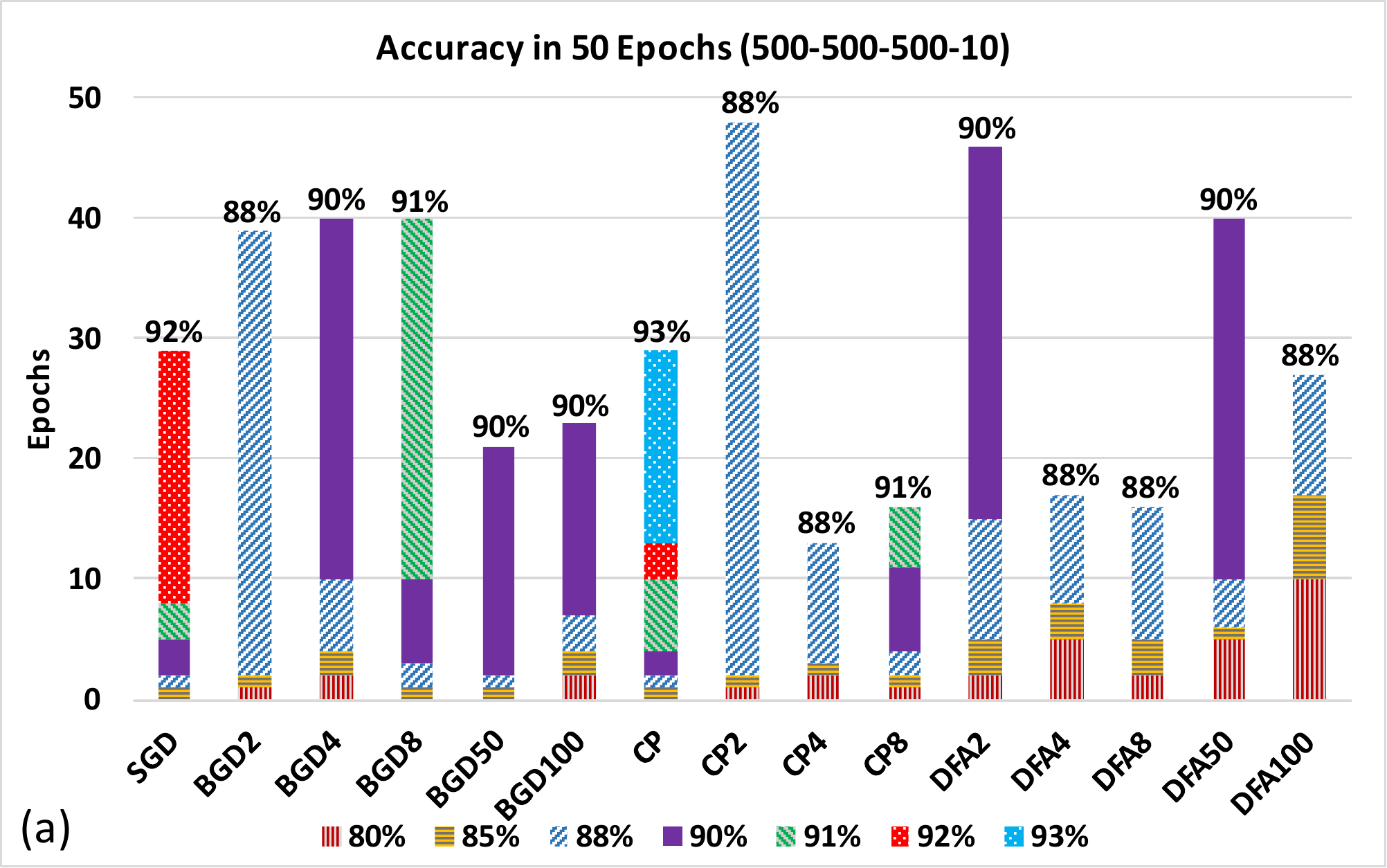} & \includegraphics[width=0.48\textwidth]{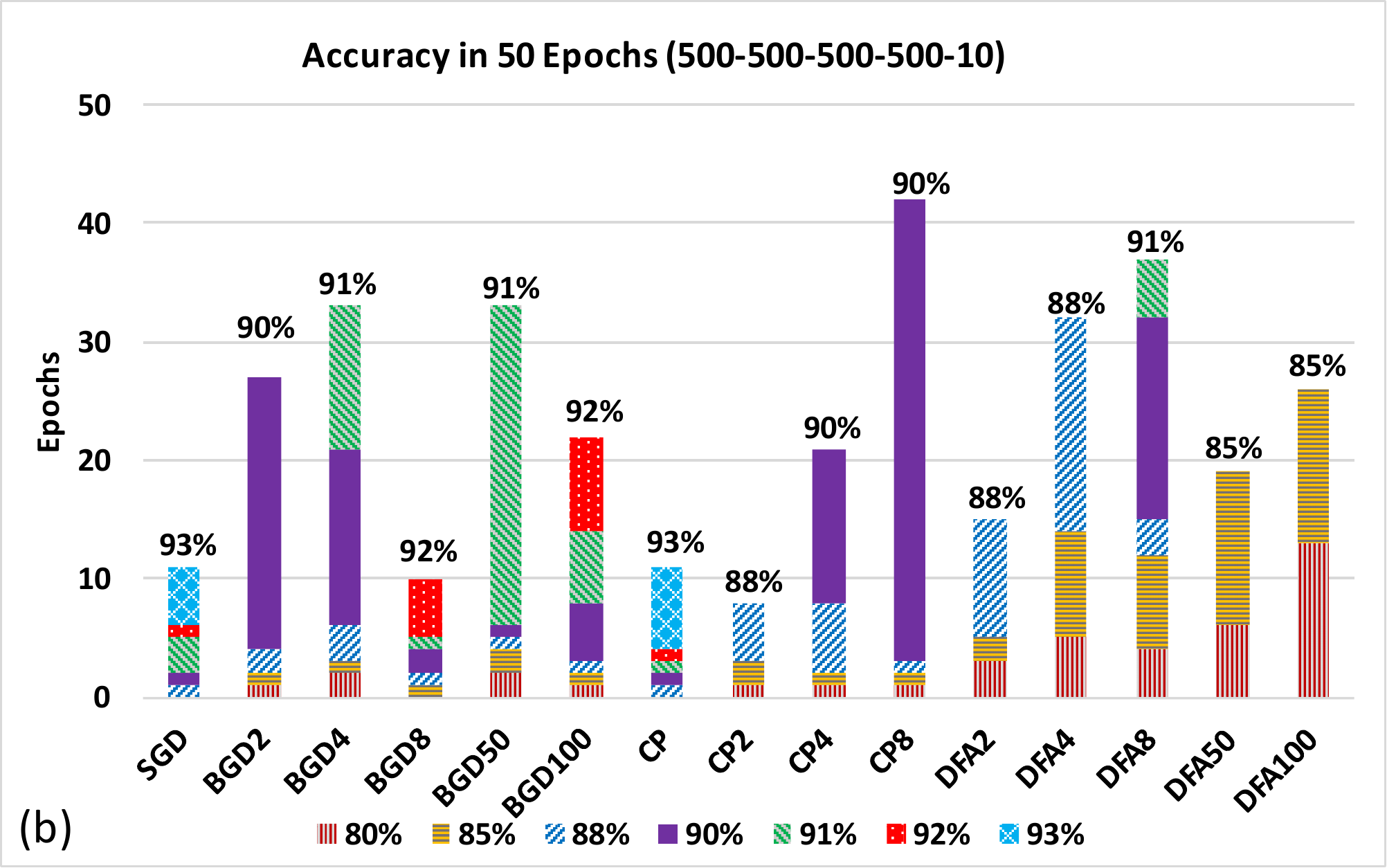}\\
\includegraphics[width=0.48\textwidth]{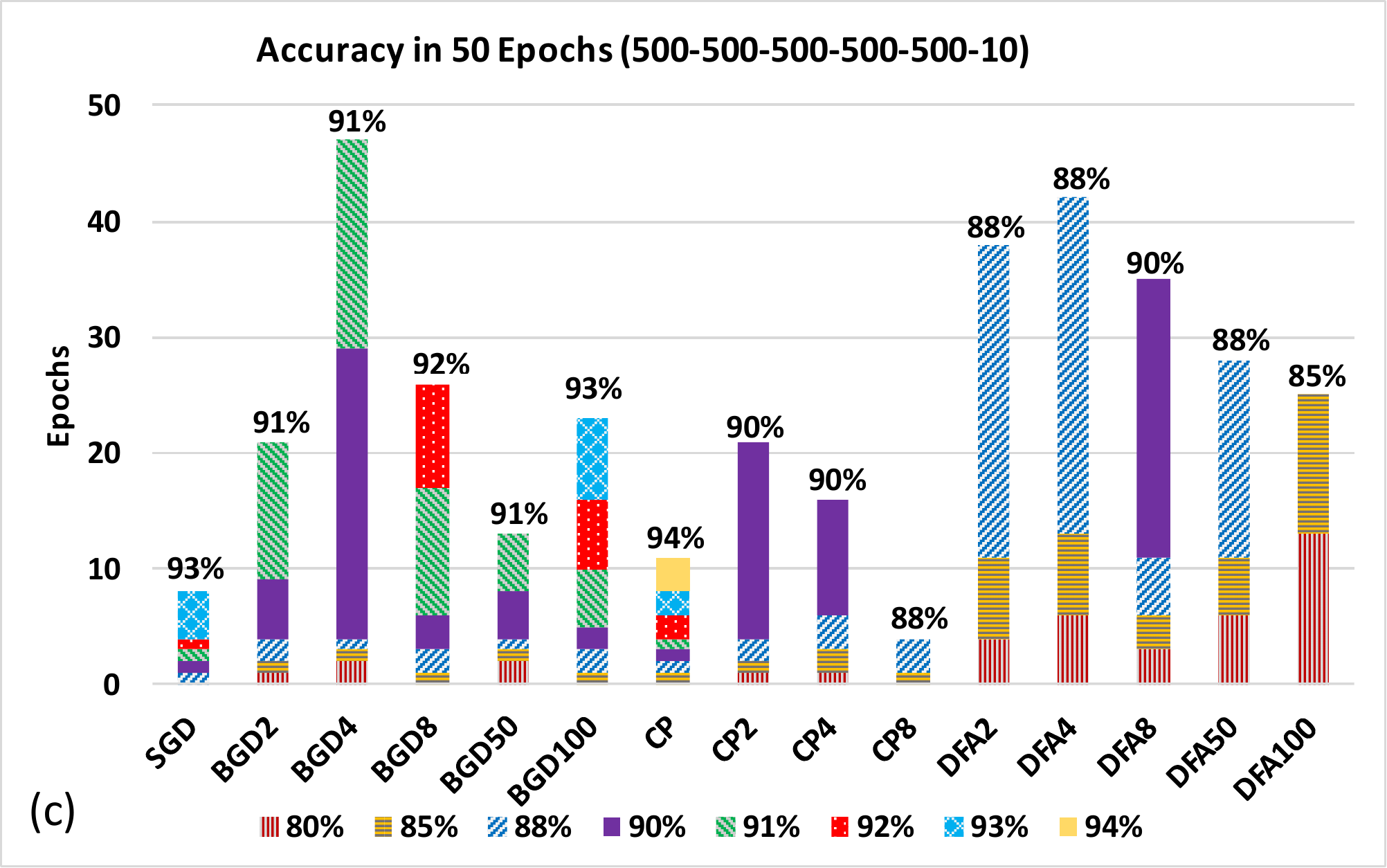} & \includegraphics[width=0.48\textwidth]{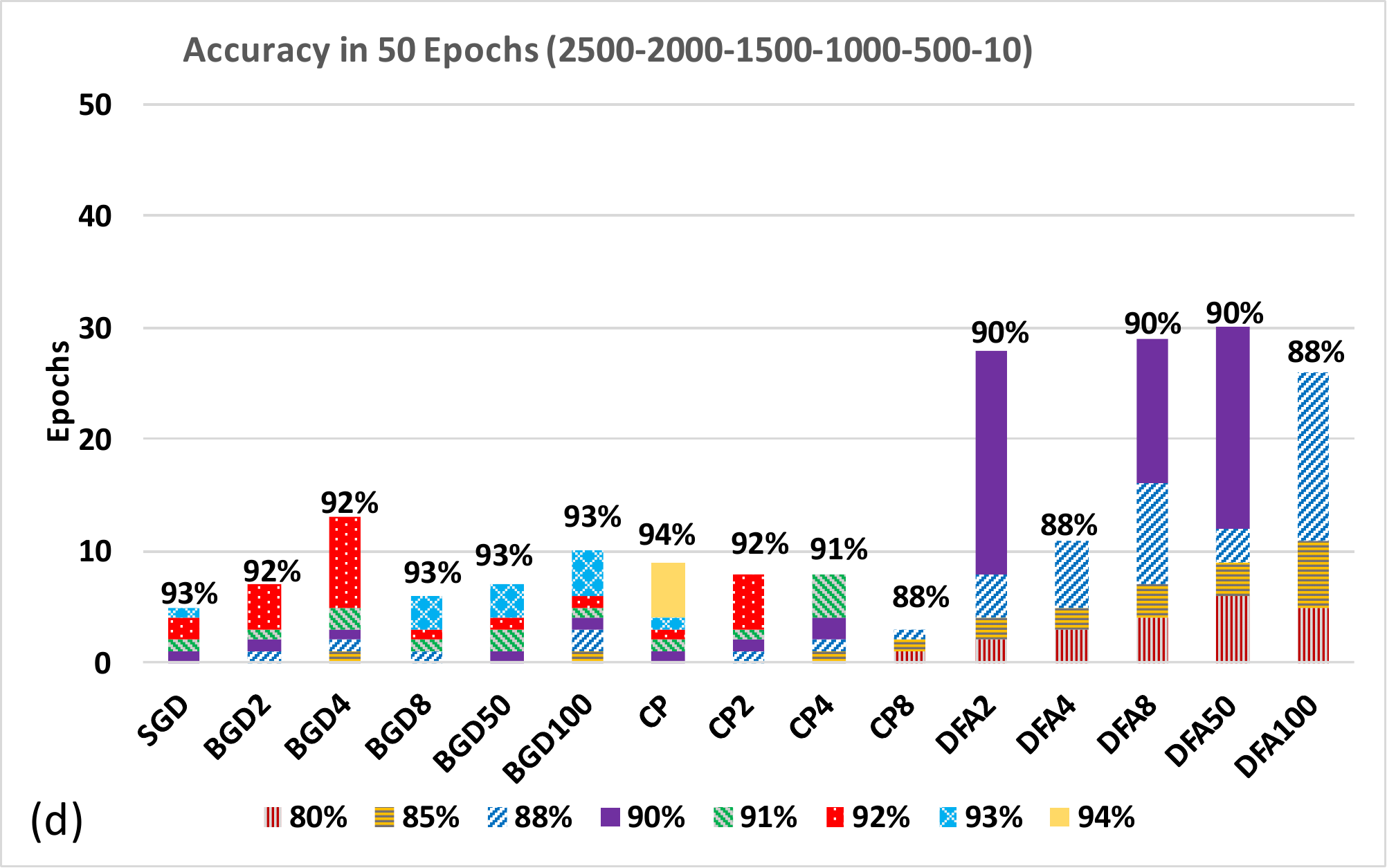}
\end{tabular}
\vspace{-10pt}
\caption{Epochs for each network to reach different accuracies for each of the training methods applied to four different neural networks.  (a) a 500-500-500-10 network trained for 50 epochs, (b) a 500-500-500-500-10 network trained for 50 epochs, (c) a 500-500-500-500-500-10 network trained for 50 epochs, (d) a 2500-2000-1500-1000-500-10 network trained for 30 epochs.}
\vspace{-15pt}
\label{fig:convergence}
\end{figure*}

\section{Evaluations} \label{sec:results}

To study the interplay between algorithms and architecture we perform two classes of studies. First, we study the convergence rate of various methods compared to each other. Next, we investigate how this rate is translated in the architecture mapping and how existing parallelism and locality affect the energy and speed to convergence. This also allows for evaluation of the proposed architecture and its various characteristics such as memory size, number of PEs per core, memory per PE, and number of cores with regard to various learning approaches.

\subsection{Methodology}

\textbf{Networks, Dataset, and Algorithms:}
We explore different network sizes and learning methods tested on a subset of the MNIST dataset in order to determine convergence and accuracy results.  All networks use ReLU as the hidden layer activation.  As the networks are trained on only a subset of the complete MNIST dataset, the accuracies achieved are lower. However, experimentation with the complete dataset and comparisons with existing results show that the relative rates of convergence and accuracies achieved by the different networks and learning algorithms behave similarly for the complete dataset.  As the purpose of this study is to compare different networks and learning methods and not to achieve the best possible accuracy, the difference between results for the complete dataset and the subset are negligible.

Four different sized networks are trained using SGD, CP, MBGD and DFA with batch sizes of 2, 4, 8, 50 and 100 for each.  To demonstrate the effect of increasing network depth, 4, 5, and 6 layer networks with hidden layers of size $500\times500$ are chosen.  A 2500-2000-1500-1000-500-10 network is used to represent a network that is both deep and wide and with varying hidden layer dimensions.

\begin{table}[!t]
\centering
\footnotesize
\vspace{-10pt}
\caption{Energy and area for FPU and SRAM blocks.}
\vspace{-5pt}
\label{table:op_vals}
\begin{tabular}{|c|c|c|}
\hline
 & Energy/Op & Area\\
\hline
Half-precision FPU & 2.63pJ & 0.0056$mm^2$\\ \hline
16KB Local SRAM & 3.5pJ (per 2 bytes) & 0.0617$mm^2$\\ \hline
512KB Off-Core SRAM & 16pJ (per 2 bytes) & $1.948mm^2$ \\ \hline
\end{tabular}
\vspace{5pt}

\vspace{-20pt}
\end{table}

\textbf{Architecture:}
Software studies showed no discernible difference between training with 16bit floating point and 32bit floating point, thus we choose to use half-precision Floating Point Units (FPUs) in the PEs.  Each PE has 16KB of local memory~\cite{lac} and each core has 512~KB of private SPAD memory.  Table~\ref{table:op_vals} shows the energy per operation for the FPU and energy per access for the local and off-core memory, as well as respective areas of the units.  Energy and area values for memories, wires, and look-up tables were obtained and estimated from \cite{dark} and CACTI~\cite{Cacti:6.0} respectively. The Half-Precision FPU area and energy were obtained from~\cite{galal2013fpu}. All estimates are for implementation in bulk CMOS operating at 1~GHz frequency.  These values are used to analytically derive time, energy and performance results for the proposed architecture.

We consider two arrangements of PEs:  $2\times 16$ cores with $16\times16$ PEs each, and $2\times 4$ cores with $4\times4$ PEs each, resulting in a total area of 103.2$mm^2$ and $178.9mm^2$ respectively. We choose to use energy required for convergence to given accuracies as the comparison unit, because of the need for a uniform measure between all networks and algorithms.

\begin{figure*}
\centering
\vspace{-25pt}
\includegraphics[width=\textwidth]{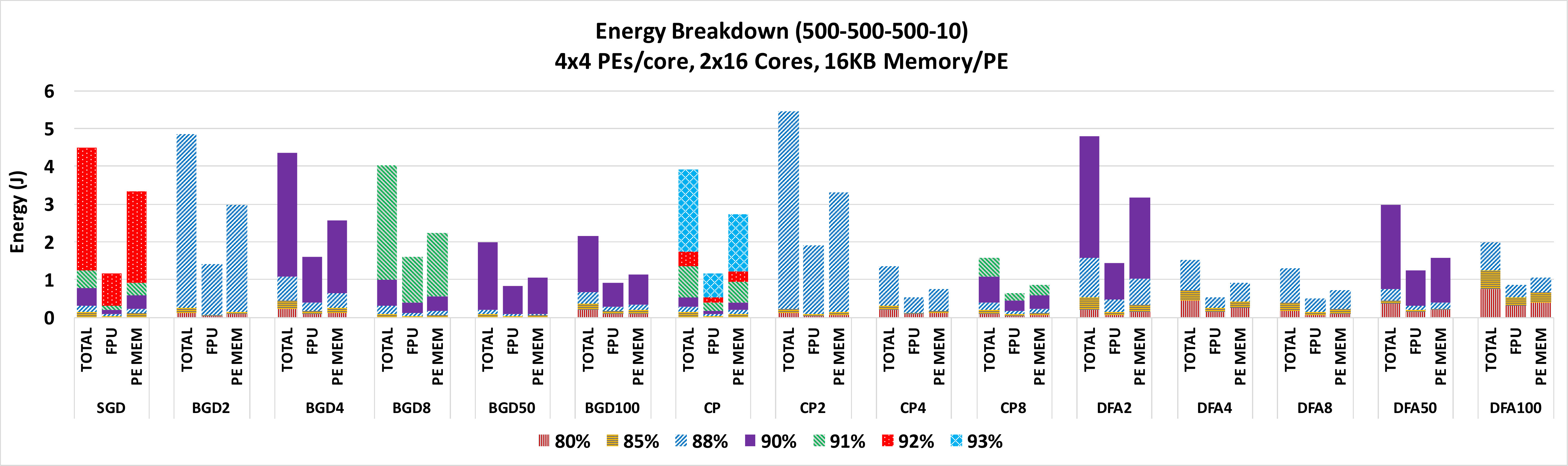}
\vspace{-10pt}
\caption{Energy required to achieve accuracy for various learning methods on a small 500-500-500-10 network that fits on $2\times16$ cores with $4\times4$ PEs each.}
\label{fig:energy_mlp1}
\vspace{-10pt}
\end{figure*}

\begin{figure*}
\centering
\includegraphics[width=\textwidth]{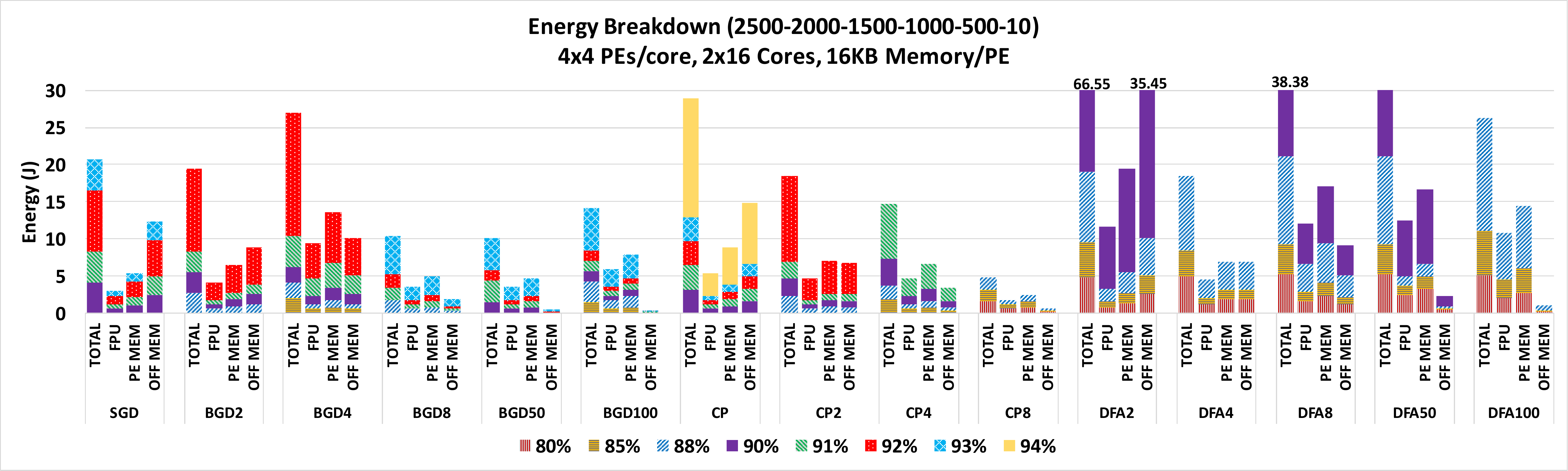}
\vspace{-20pt}
\caption{Energy required to achieve accuracy for various learning methods on a large 2500-2000-1500-1000-500-10 network that does not fit in local PE memory.  High communication cost is required for bringing in weights from off-core memory for SGD and CP; these two algorithms would not be used in practice but energy results are presented here for completeness}
\vspace{-10pt}
\label{fig:energy_mlpdiff}
\end{figure*}

\begin{figure*}
\centering
\includegraphics[width=\textwidth]{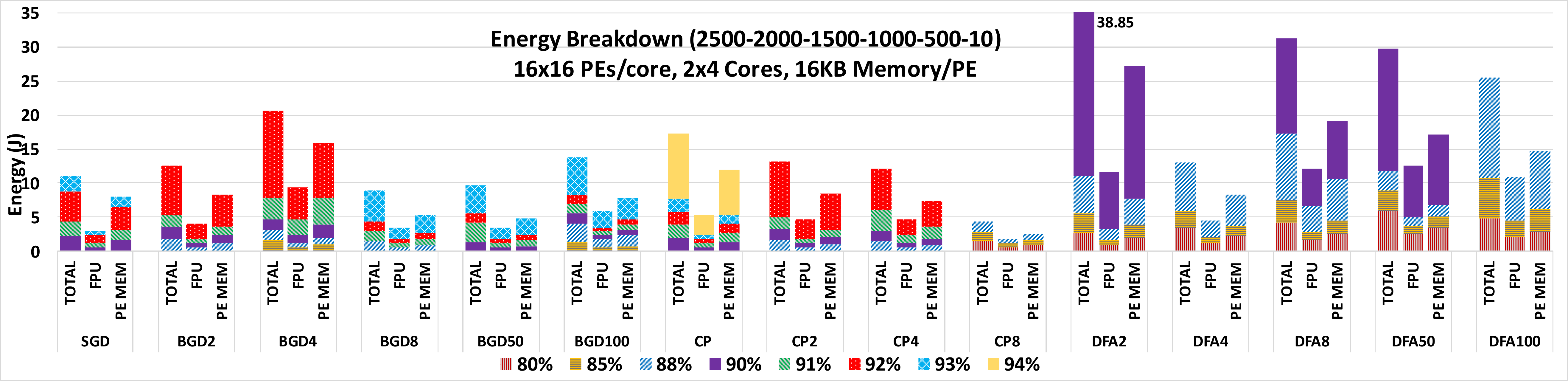}
\vspace{-10pt}
\caption{Energy required to achieve accuracy for various learning methods on a large 2500-2000-1500-1000-500-10 network that fits on $2\times4$ cores with $16\times16$ PEs each.}
\vspace{-15pt}
\label{fig:energy_mlp3}
\end{figure*}

\subsection{Software Experimental Results}

Figure~\ref{fig:convergence} shows the validation accuracy achieved for the four chosen networks.  Each epoch constitutes a single pass over the entire dataset.  The network in Figure~\ref{fig:convergence}(a) is small enough that even SGD and CP require many epochs to reach convergence, although the accuracy reached is higher than for other algorithms.  In Figure~\ref{fig:convergence}(b), the additional hidden layer causes the epochs to convergence for SGD and CP to drop by 60\% so that they converge faster and to a higher accuracy than the minibatched algorithms.  In general, SGD and CP are able to achieve the highest accuracy in the fewest epochs of all algorithms as the weights are updated once for each sample in an epoch.  CP also performs as well or better than SGD in all cases, although this is not true for the MBCP.  Further for small minibatches, minibatches of size eight outperform those of size two and four for all networks as it can support a slightly higher learning rate.

For the same learning method compared across the different networks, larger networks are able to converge to higher accuracies in fewer epochs.  This behavior is especially evident between the network in Figure~\ref{fig:convergence}(c) and the largest network in Figure~\ref{fig:convergence}(d).  Here, instead of increasing depth of the network, the size of each layer is increased.  However, the larger size of the network means more aggregate calculations and weight updates, which will have an impact on the energy and time performance when mapped to the architecture.

The difference in convergence rate and highest achievable accuracy between learning methods also becomes less evident as the network size increases.  Thus as network size increases, MBGD's performance approaches that of SGD in terms of accuracy and is also able to reach this accuracy in a comparable number of epochs.  The architectural implication is that minibatch training can be used for larger networks that do not fit on local memory without sacrificing accuracy. Note that in the same number of epochs, DFA always achieves a lower accuracy than other learning methods due to the lower learning rate.

\subsection{Architecture Experimental Results}

Figures~\ref{fig:energy_mlp1}-\ref{fig:energy_mlp3} demonstrate the energy required for three networks as well as the breakdown into FPU energy and memory access energy.  The energy of broadcasts was found to be negligible and is not included here.  Network 1 in Figure~\ref{fig:energy_mlp1} is small and fits completely on the local core memories for all configurations.  For the same 90\% accuracy, SGD requires 70\% of the energy as MBGD for large minibatches while CP requires 30\%.  Examination of energy breakdowns shows that for minibatch algorithms, energy usage is dominated by the FPU while for SGD memory access energy is 1.5 times higher than FPU energy.  The energy cost for CP is split evenly between FPU and memory accesses.  This is due to the fact that both minibatched and CP algorithms reduce the number of weight accesses required each epoch compared to SGD.

\begin{figure*}[!t]
\centering
\vspace{-20pt}
\includegraphics[width=\textwidth]{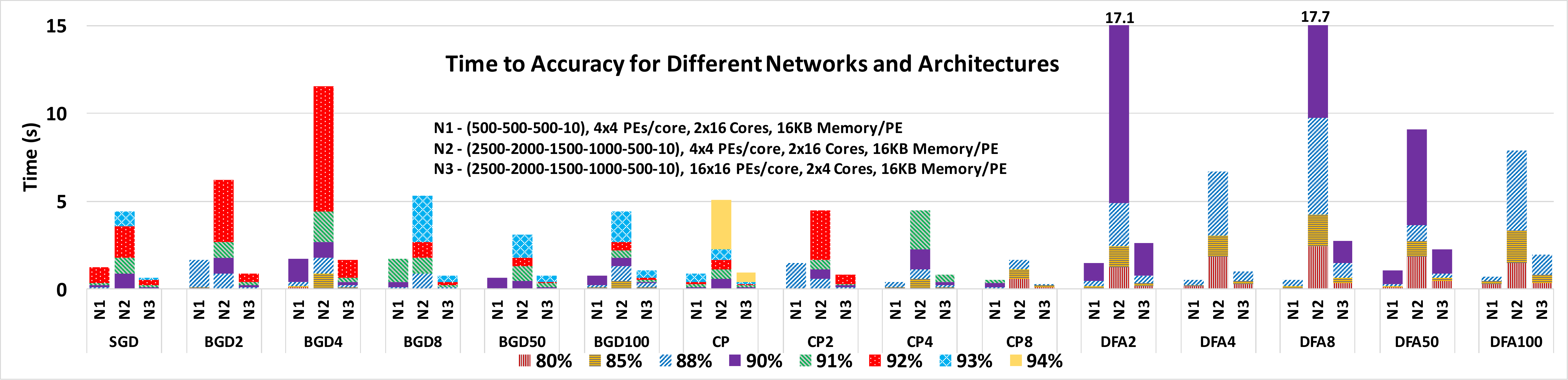}
\vspace{-20pt}
\caption{Time required to achieve accuracy for various learning methods and networks.}
\label{fig:time}
\end{figure*}

Network 2 in Figure~\ref{fig:energy_mlpdiff} is representative of a large network that does not fit in local core memories.  Although the minibatch algorithms require more epochs to converge to the same accuracy as SGD and CP, their total energy consumption is lower due to the smaller number of weight accesses.  For networks that do not fit on the core, SGD and CP must access weights from off-core, substantially increasing energy usage.  As discussed previously, SGD and CP also require higher bandwidth to access the weights if they do not fit locally.  These results suggest to perform training using minibatch algorithms for networks that do not fit.

Network 3 in Figure~\ref{fig:energy_mlp3} is the same as network 2 but trained on an architecture with more PEs such that the network fits locally.  Unlike the smaller network 1, MBGD with batch size of 50 now performs better than SGD in terms of energy cost, although still not as well as CP.  Comparison of FPU energy only shows that batched algorithms have a higher energy cost due to the greater number of epochs, but this is balanced out by the lower memory access energy.  As discussed in the previous section, MBGD's performance in terms of epochs increases with network size.  Here, the difference in energy consumption for FPU operation between minibatched and non-minibatched methods is small enough that memory access energy becomes the dominant factor differentiating the two methods.  However, the faster convergence and higher accuracy of CP causes it to perform better than MBGD even for large networks.

These results suggest that for large networks, MBGD can perform better in terms of energy than SGD even when there is enough local memory to store the entire network.  Further, CP consistently outperforms all other training methods.  The energy to convergence results for DFA indicate that although direct propagation of the error from the last layer leads to fewer FPU operations, thus less energy required per epoch, the slower convergence rate causes the algorithm to consume more energy to reach the same accuracy.

\begin{figure}[!t]
\vspace{-20pt}
\centering
\includegraphics[width=0.45\textwidth]{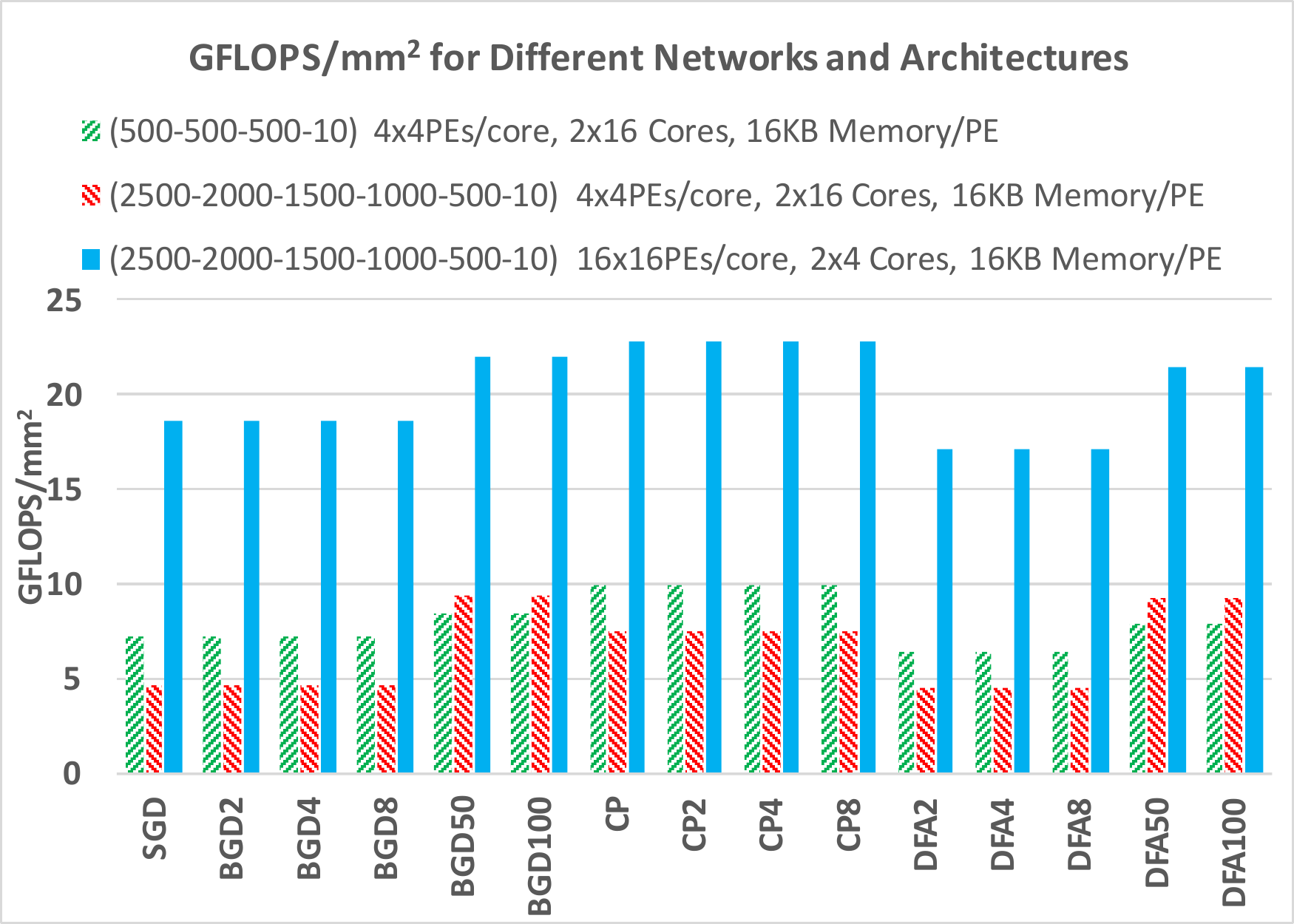}
\vspace{-10pt}
\caption{GFLOPS/$mm^2$ for various learning methods and networks.}
\vspace{-15pt}
\label{fig:gflop}
\end{figure}

Figure~\ref{fig:time} illustrates the time required for each network to reach given accuracies on the architecture operating at 1GHz.  As in the energy performance comparisons, CP performs better than SGD and requires less time to converge to the same accuracy.  While software studies showed that the epochs required to converge to the same accuracy was similar for the two algorithms, the pipelining and weight locality utilized by CP allows it to achieve better performance on the architecture.  When the network does not fit in local memory for CP and SGD, the need to fetch weights from off-core memory can significantly increase the time to convergence.  For Network 2 which does not fit on the core, the utilization drops from 99\% to 75\% for CP and from 81\% to 47\% for SGD.  From software studies, the epochs to convergence for minibatched algorithms is similar to batched for large network sizes.  Architecturally, however, minibatched algorithms can converge faster than non-minibatched algorithms if the network does not fit even when it requires more epochs.

\begin{table}[!t]
\vspace{-15pt}
\centering
\caption{GFLOPS/W for (a) a small network that fits on $2\times 16$ cores of $4\times 4$ PEs; (b) a large neural network that does not fit on the same architecture; (c) the same neural network that fits on 
$2\times 4$ cores of $16\times16$ PEs.}
\vspace{-10pt}
\scriptsize
\begin{tabular}{|c|c|c|}
\hline
Network Dimensions& BP Method & GFLOPS/W\\
\hline
 & SGD & 177\\
500-500-500-10 & CP & 204\\
 & MBGD & 195\\
 \hline
 & SGD & 98\\
2500-2000-1500-1000-500-10 & CP & 127\\
 & MBGD & 187\\
\hline
 & SGD & 185\\
2500-2000-1500-1000-500-10 & CP & 211\\
 & MBGD & 195 \\ \hline
\end{tabular}

\vspace{-15pt}
\label{table:performance}
\end{table}

Figure~\ref{fig:gflop} and Table~\ref{table:performance} show the performance in GFLOPS/W and GFLOPS/$mm^2$ for SGD, CP and MBGD applied to networks of different sizes respectively.  CP consistently outperforms SGD in all cases.  For networks that do not fit on core, MBGD demonstrates the highest performance, followed by CP and SGD, while for networks that do fit, CP outperforms MBGD.  Although performance of MBGD can be greater than CP/SGD due to the reduced energy cost of accessing fewer weights, it is not as accurate as either, especially for small networks (90\% vs. 92\% accuracy for small networks and 93\% vs 94\% accuracy for large networks) and also takes longer to converge. Thus, there is a tradeoff between performance and accuracy/time to convergence that must be considered when determining which training method to use.  
When comparing between networks of different sizes, it can be seen that for the same architecture size, while the larger network reaches higher accuracy, the time to convergence to a lower accuracy is smaller for smaller networks.  The flexibility of the architecture in supporting both batched and non-batched training algorithms provides the user with freedom to determine the learning method to use based on the time and accuracy constraints of their network application.  

The best overall performance of the architecture occurs for a network of size 2500-2000-1500-1000-500-10 mapped to a $2\times4$ array of cores with $16\times16$ PEs, with training done using CP.  From an algorithmic perspective, the size and depth of the network leads to higher accuracy and fewer epochs to convergence while the greater number of PEs both increases the effective utilization of the architecture and eliminates costly accesses to external memory.  

CATERPILLAR achieves 98\% effective utilization of the FPUs and performance of 211~GFLOPS/W for networks that fit on cores using CP. Further, when the same network does not fit on the cores, using minibatched algorithms MBGD can achieve 187~GFLOPS/W at 94\% utilization.

\section{Related Work}\label{sec:related}
Several FPGA implementation efforts have been performed to accelerate the training of neural networks ~\cite{park2016fpga}~\cite{fpga}~\cite{processor}.  Maximum performance of up to 10 G Multiply Accumulates (GMACs) is achieved in~\cite{processor}.  However, these works are limited in scope as they focus on either retraining~\cite{park2016fpga} or shallow 2-layer neural networks~\cite{fpga}.  Furthermore, there is no support for performing different learning algorithms as for Caterpillar architecture.

 Gupta et al. have shown that 16-bit fixed precision can achieve the same accuracy as floating-point if stochastic rounding is used~\cite{stochastic}. Our preliminary studies suggests that the convergence rate and accuracy decreases for networks deeper than two layers with stochastic rounding. Further study is required to completely characterize the performance of stochastic rounding compared to floating point.

The work in~\cite{cp} showed that CP can outperform MBGD's speed and accuracy for CNNs. In this paper we apply and evaluate CP for MLPs.

\section{Conclusion} \label{sec:conclusion}

Our investigation for training MLPs demonstrates that for various networks sizes, the target architecture should support both GEMV (for pipelined backpropagation), GEMM (for minibatched algorithms), and hierarchical collective communications.  For networks that do not fit on chip, minibatched algorithms have comparable performance to pipelined backpropagation, however for networks that fit, pipelined backpropagation consistently performs the best. 
Fast convergence  on  the  algorithmic  side  in  tandem  with  layer parallelization and weight locality from an architectural perspective allows Pipelined Continuous Propagation to outperform all other training methods in terms of energy and time to convergence, distinguishing it as  a  promising  training  method for use with specialized deep learning architectures.

\section*{Acknowledgments}
We thank Hadi Esmaeilzadeh, Michael James, David Koeplinger, Ilya Sharapov, Vijay Korthikanti, and Sara O'Connell for their feedback on the manuscript.
This research was partially sponsored by NSF grants
CCF-1563113.
Any opinions, findings and conclusions or recommendations expressed in this material are those of the authors and do not necessarily reflect the views of the National Science Foundation (NSF).



\bibliographystyle{IEEEtran}
\bibliography{references2}

%

\end{document}